\documentclass[12pt]{article}
\usepackage{graphicx}

\def\nikhef{NIKHEF, Amsterdam, The Netherlands}

\def\Title#1{\begin{center} {\Large #1 } \end{center}}
\def\Author#1{\begin{center}{ \sc #1} \end{center}}
\def\Address#1{\begin{center}{ \it #1} \end{center}}

\def\Bsmumu{$B_s \rightarrow \mu^+ \mu^-$}

\newenvironment{Abstract}{\begin{quotation}  }{\end{quotation}}
\newenvironment{Presented}{\begin{quotation} \begin{center} 
             PRESENTED AT\end{center}\bigskip 
      \begin{center}\begin{large}}{\end{large}\end{center} \end{quotation}}
\def\Acknowledgements{\bigskip  \bigskip \begin{center} \begin{large}
             \bf ACKNOWLEDGEMENTS \end{large}\end{center}}

\def\beq{\begin{equation}}
\def\eeq#1{\label{#1}\end{equation}}
\def\eeqn{\end{equation}}

\def\beqa{\begin{eqnarray}}
\def\eeqa#1{\label{#1}\end{eqnarray}}
\def\eeqan{\end{eqnarray}}

\let\bar=\overbar

\def\Dslash{\not{\hbox{\kern-4pt $D$}}}
\def\dslash{\not{\hbox{\kern-2pt $\del$}}}

\def\msb{{\bar{\ssstyle M \kern -1pt S}}}

\begin{document}
\begin{titlepage}

\vfill
\Title{$B_s\rightarrow \mu^{+}\mu^{-}$ the at LHC}
\vfill
\Author{ Nicola Serra}
\Address{\nikhef}
\Author{On behalf of the LHCb, ATLAS and CMS collaborations}
\vfill
\begin{Abstract}
The ATLAS, CMS and LHCb experiments will perform extensive searches for physics
Beyond the Standard Model (BSM). 
The investigation of decays of beauty hadrons represents an alternative and complementary approach to the direct BSM
searches. \\ 
A particularly promising observable for the search of New Physics (NP) in $B-$hadron decays, 
is the measurement  of the branching ratio of the decay $B_S\rightarrow \mu^{+} \mu^{-}$. 
This observable is sensitive to physics BSM with new scalar or pseudoscalar 
effective operators, such as theories involving an extended Higgs sector. 
Here the prospects of the ATLAS, CMS and the LHCb experiments for such a measurement are discussed. 
In particular the LHCb experiment, thanks to its good particle identification and 
momentum resolution, has the potential for an early discovery of this decay.  
\end{Abstract}
\vfill
\begin{Presented}
CKM2010 - $6^{th}$ International Workshop on the CKM Unitarity Triangle\\

\end{Presented}
\vfill
\end{titlepage}
\def\thefootnote{\fnsymbol{footnote}}
\setcounter{footnote}{0}

\section{Introduction}

The decay of beauty mesons is a promising strategy for the indirect search of NP. 
Among these decays one of the most prominent is the rare decay $B_s\rightarrow \mu^+ \mu^-$. 
Its branching ratio is precisely predicted within the SM: $BR(B_s\rightarrow \mu^+\mu^-)=(3.6\pm0.4)\times 10^{-9}$~\cite{burasBsmumu}. 
This decay is both a Flavor Changing Neutral Current (FCNC) and helicity suppressed. 
Its general effective hamiltonian can be written as:
\begin{equation}
H = -2\sqrt(2)G_F |V_{tb}^{*}V_{ts}| \left( C_P O_P + C_S O_S + C_{10} O_{10} \right)
\end{equation}
where $V_{ij}$ are elements of the CKM matrix, the $O_i$ are effective operators of the OPE expansion,  
the $C_i$ are the Wilson coefficients and $G_F$ is the Fermi constant.
%
In the SM the contribution of the coefficients $C_P$ and $C_S$ is negligible. 
However, in NP  with new effective scalar operators, such as 
theories involving an extended Higgs sector, the contribution of such coefficients  can lift the helicity suppression, 
enhancing the branching ratio of the $B_s\rightarrow \mu^+ \mu^-$ decay. 
Well known examples are theories with two Higgs doublets, such as the 2HDM or the 
popular Minimal Supersymmetric extension of the Standard Model (MSSM). 
In the 2HDM the branching ratio of the $B_s \rightarrow \mu^+ \mu^-$ is proportional to the fourth power 
of the ratio of the vacuum expectation value of the two doublets ($\tan \beta $), while in the MSSM this observable is further enhanced for high values of $\tan\beta$, 
according to the formula:
\begin{equation}
BR(B_s\rightarrow \mu^{+}\mu^{-})\propto \frac{\tan^6\beta}{M_A}, 
\end{equation}
where $M_A$ is the mass of the $CP$-odd neutral Higgs.
It is worth noticing that even in the presence of Minimal Flavor Violating (MFV) theories, where the flavor structure is
described by the CKM matrix, this branching ratio could still be highly enhanced.   
The present upper limit for this decay is $BR(B_s\rightarrow \mu^+\mu^-)< 4.3 \times 10^{-8}$ at $95\%CL$, set by the CDF Collaboration~\cite{CDFBsmumu} 
and $BR(B_s\rightarrow \mu^+\mu^-)< 5.1\times 10^{-8}$ at $95\%CL$, set by the D0 Collaboration~\cite{D0Bsmumu}. 
These limits are still one order of magnitude higher than the SM prediction and therefore there is still a large scope for NP contributions.

\section{Event Selection}
\begin{figure}[!b]
\begin{center}
\includegraphics[width=0.33\textwidth,angle=270]{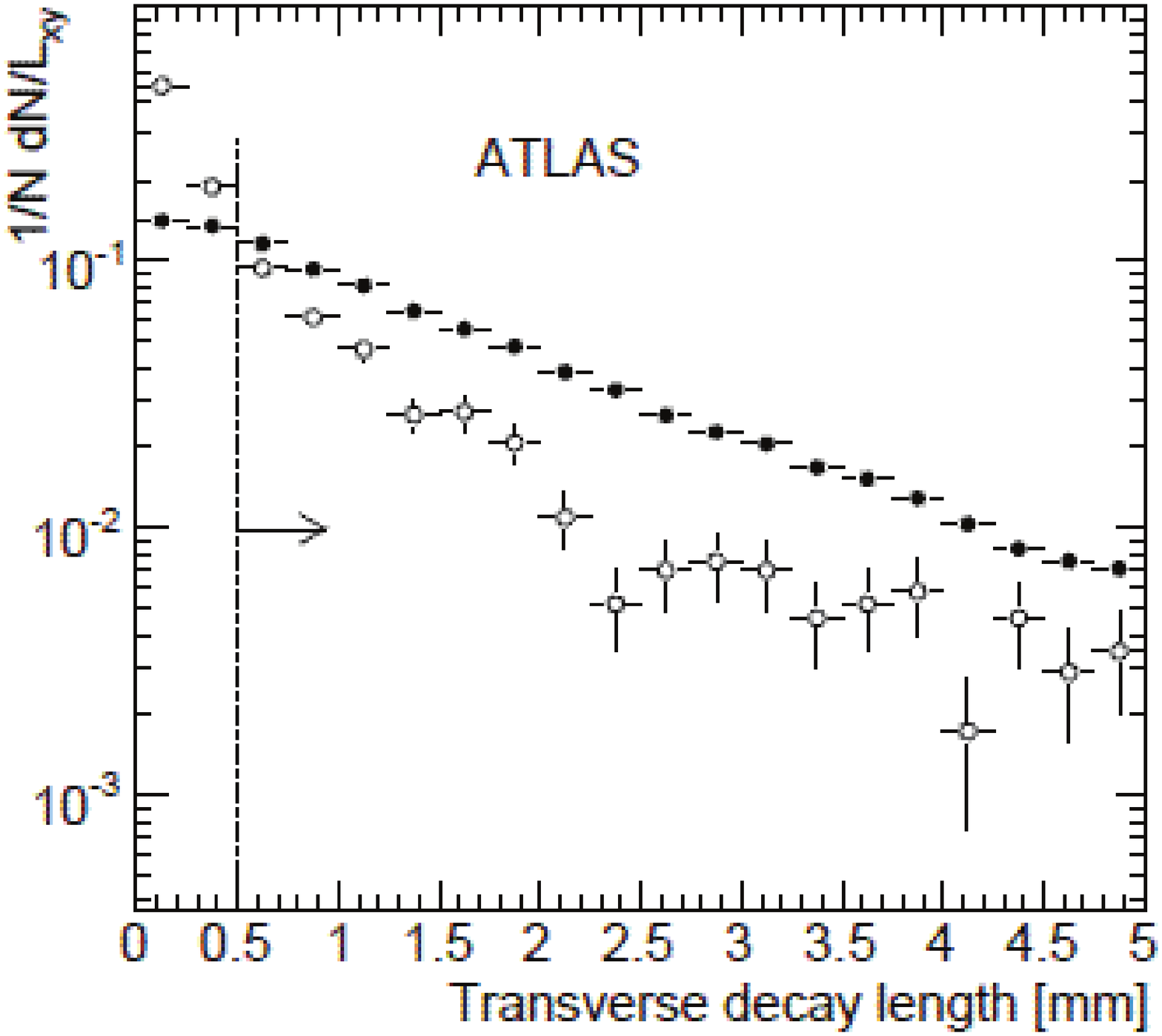}
\includegraphics[width=0.33\textwidth,angle=270]{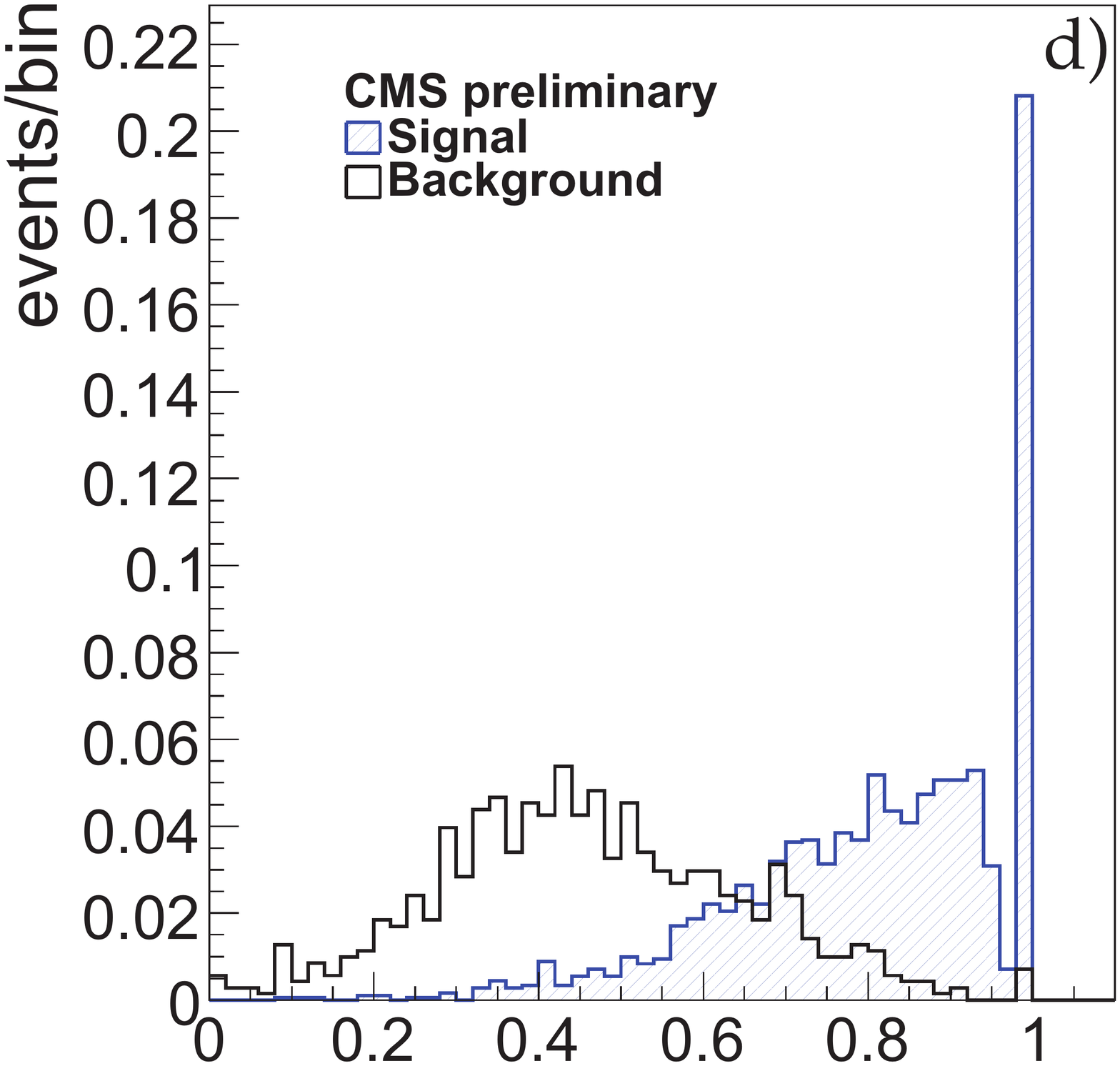}
\caption{Distribution of two discriminating variables, for the ATLAS and CMS experiments. The two variables are the 
transverse decay length of the $B_S$ (ATLAS) and the isolation of the $B_s$ (CMS).}\label{Fig:variables}
\end{center}
\end{figure}

The three experiments ATLAS~\cite{ATLAS}, CMS~\cite{CMS} and LHCb~\cite{LHCb} at the LHC accelerator all have a 
$B-$physics program  which includes the search for the rare decay \Bsmumu . 
ATLAS and CMS are general purpose experiments, 
covering the central rapidity region $|\eta |<2.5$. Their trigger for $B-$hadrons relies on muons with high transverse momentum ($p_T>(4-6)$GeV). 
The LHCb experiment is devoted to the study of $B-$hadrons. 
The detector consists of a single arm forward spectrometer, covering the region $1.9<\eta <4.9$, 
which allows lower $p_T$ muons to be triggered than in the case of the central detectors. 
This results in an increase of the effective $b\overline{b}$ cross section, by about a factor $2.3$. \\
Since the branching ratio of the $B_s\rightarrow \mu^{+} \mu^{-}$ is very small, it is important to efficiently select the 
signal while rejecting the vast majority of background events. 
The main background source consists of combinatorial double semi-leptonic decays $b\overline{b}\rightarrow \mu^+ \mu^-$, 
where the two muons accidentally form a secondary vertex.
Other background sources (which, in the case of LHCb, give a negligible contribution after the selection requirements) are 
the two body charmless decays $B_{d,s} \rightarrow h^+ h^-$, where $h = K,\pi$ is misidentified as a muon, and 
decays such as $B \rightarrow J/\phi h X$, where $h$ is either a muon or misidentified as a muon. \\
The selection strategy for the ATLAS and CMS experiments uses a cut based selection~\cite{ATLASBsmumu}, \cite{CMSBsmumu}. The discriminating variables are 
the flight distance of the $B_s$, the quality of the reconstructed secondary vertex, the consistency between the $B_s$ 
flight direction and its momentum (pointing angle) and the isolation of the $B_s$ (i.e. requiring that
 the $B_{s}$ has few tracks in a cone around its flight direction). 
Some discriminating variables are shown in Fig.\ref{Fig:variables}. 

\begin{figure}[!b]
\begin{center}
\includegraphics[width=0.5\textwidth,angle=0]{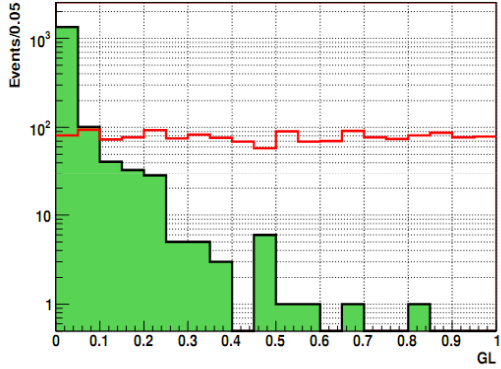}\label{GL}
\caption{Geometrical likelihood distribution, for signal and background, according to the Monte Carlo simulation. 
The open histogram is the signal, while the filled histogram is the background. The background corresponds to about 5pb$^{-1}$.}
\end{center}
\end{figure}

The LHCb analysis strategy is based on a ``soft''  selection, in order to reject most of the background keeping 
a high signal efficiency, followed by a likelihood analysis~\cite{LHCbRoadmap}. 
A geometrical likelihood is built using several geometrical discriminating variables of the event. 
In Fig.~\ref{GL} the expected geometrical likelihood distribution is shown, for signal and background, according 
to the Monte Carlo simulation.

Fig.3 (right) shows the comparison between data and Monte Carlo simulation for the invariant mass of di-muon events, the geometrical likelihood after selection is shown in Fig.3 (left).
These and other preliminary studies show that the LHCb expectation, based on simulation, agrees with the early data.  
\begin{figure}[!b]
\begin{center}
\includegraphics[width=0.45\textwidth,angle=0]{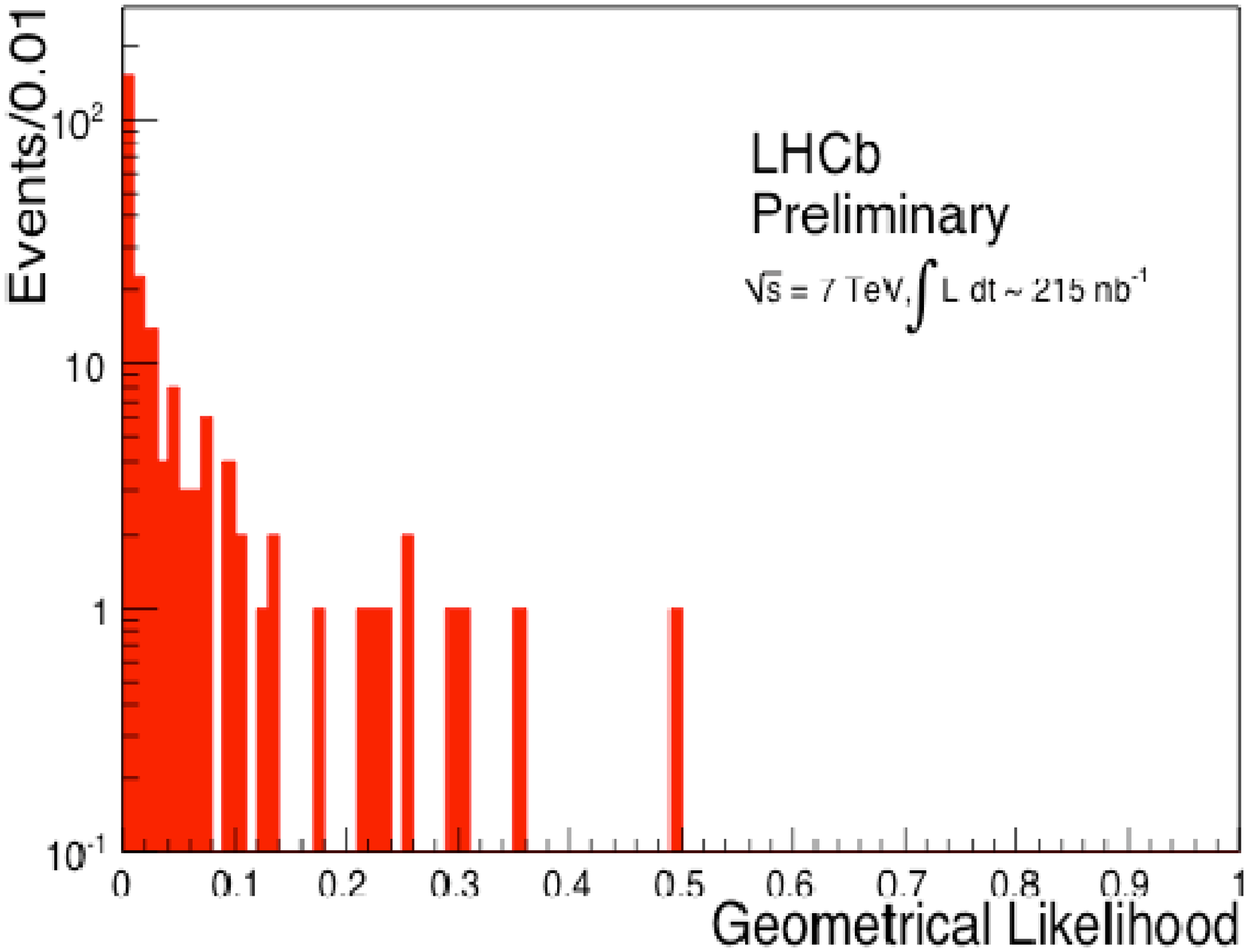}
\includegraphics[width=0.45\textwidth,angle=0]{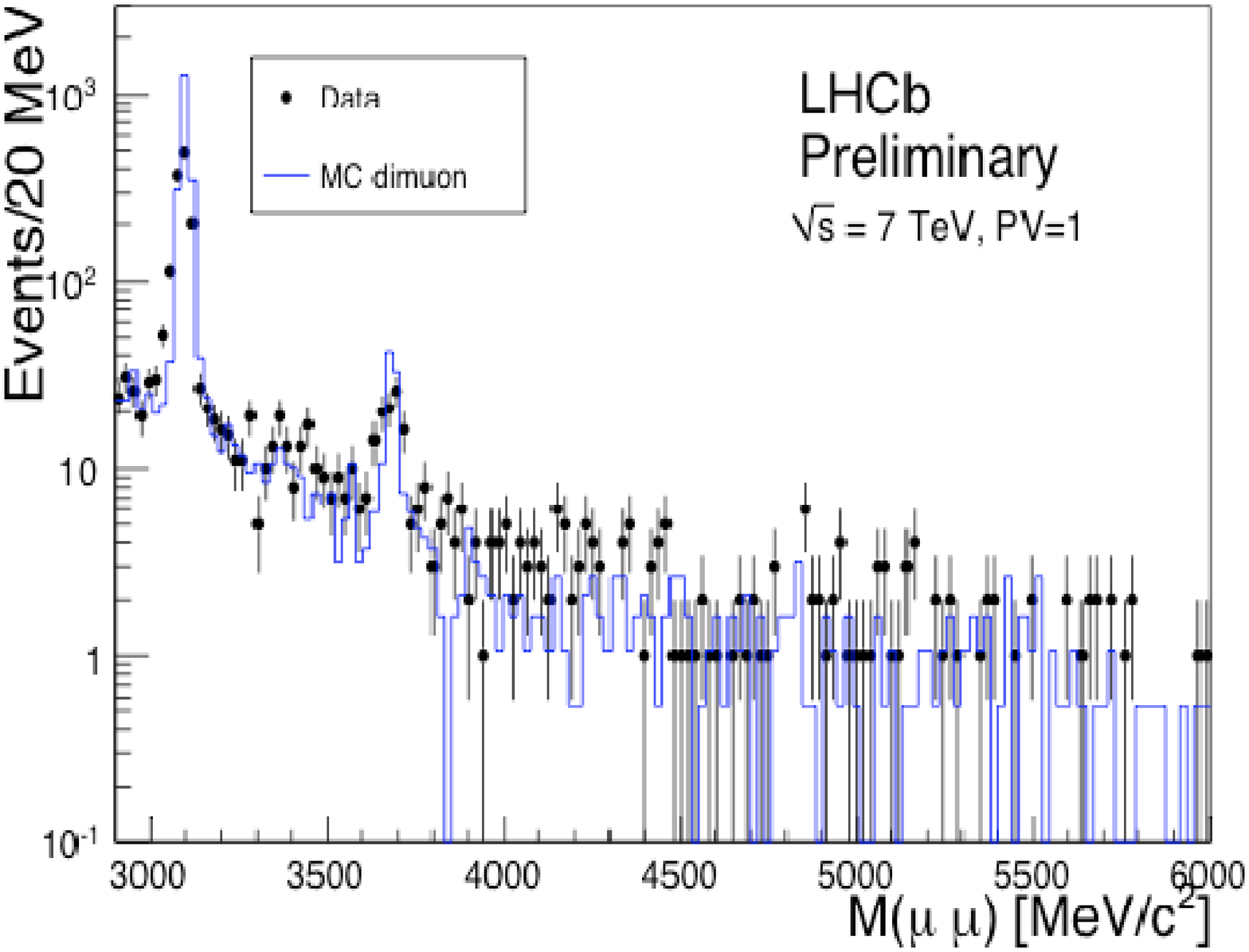}
\caption {Geometrical likelihood for the background side bands, after the ``soft'' selection (left) and 
data MC comparison for the di-muon invariant mass distribution (right). The corresponding luminosity is about 0.2pb$^{-1}$.}
\end{center}
\end{figure} 
After the ``soft'' selection, the selected events are binned in three variables: 
the geometrical likelihood, the muon particle identification likelihood and the di-muon invariant mass distribution. 
The sensitivity of each bin is combined using the Modified Frequentist Approach (MFA)~\cite{MFA}. 

\section{Normalization}

At hadron colliders it is difficult to perform an absolute measurement of a branching ratio, since this would require precise measurements of he $b\overline{b}$ 
cross section, the luminosity and the fragmentation functions. 
All the LHC experiments will therefore make a relative measurement with respect to a known B decay, according to the 
formula: 
\begin{equation}
BR(B_s\rightarrow \mu^{+} \mu^{-})=BR_{cal}\times \frac{\epsilon}{\epsilon_{cal}}\frac{f_{cal}}{f_{s}}\frac{N}{N_{cal}}, 
\end{equation} 
where the subscript $cal$ refers to a calibration channel, $\epsilon$ is the total efficiency (which includes also the 
acceptance), $f$ is the fragmentation function and $N$ is the number of events. 
All the three experiments are studying the possibility of normalizing to the decay channel $B^+\rightarrow J/\psi K^+$. 
This channel, in addition to having a well known branching ratio,  has the advantage of having two muons 
in the final state, therefore the same trigger and the same 
particle identification likelihood. Since it is not a $B_s$ decay, 
the normalization would require knowledge of the ratio of the fragmentation functions $f_{d}/f_{s}$. 
As this is known with 15$\%$ precision~\cite{PDG}, this would form the limiting source of uncertainty for this approach. 
However the ratio $f_d/f_s$ can be measured either with 
semi-leptonic or hadronic decays~\cite{fdfsHadronic}, by using the isospin symmetry assumption $f_u=f_d$. 
Another possibility is to normalize to the decay $B_s\rightarrow J/\psi \phi$. 
In this case the ratio of fragmentation functions cancels out, however at present this branching ratio is still 
poorly known. \\
Concerning the LHCb analysis strategy, in order to use the MFA, it is important to know also the fraction of 
signal events falling into each GL bin. 
For this calibration the LHCb collaboration plans to use two body charmless decay $B \rightarrow h^+ h^-$. 
These channels have the same topology as the signal but the trigger is very different. 
This effect can be accounted for using events triggered independently of the signal, where 
other particles in the event caused the trigger~\cite{LHCbRoadmap}. 

\section{Sensitivity}
Assuming the SM branching ratio, an integrated luminosity of 10fb$^{-1}$ and a collision energy of $14$TeV, the ATLAS experiment expects 
$5.6$ signal events and $14^{+13}_{-10}$ background events from $b\overline{b}\rightarrow \mu^{+}\mu^{-}$. \\
It is interesting to compare the CMS and the LHCb expectations. 
Assuming an integrated luminosity of $1$fb$^{-1}$ at a collision energy of 7TeV, CMS expects $1.4$ signal events and 4.0 background events, 
while LHCb expects $6.3$ signal events and $32.4$ background events in the most sensitive region. 
Using the modified frequentist approach this would lead to a limit $BR(B_s\rightarrow \mu^+ \mu^-)<15.8\times 10^{-9}$ at $90\%C.L.$ for CMS and 
$BR(B_s\rightarrow \mu^+ \mu^-)<7.0\times 10^{-9}$ at $90\%C.L.$ for LHCb.
It should be noted that this comparison is made using the same statistical method (MFA) for CMS and LHCb, using as input the 
expected number of background and signal events scaled for 7TeV. 
The CMS Collaboration quotes a limit of $BR(B_s\rightarrow \mu^+\mu^-)<16.0\times 10^{-9}$ 
for an interaction energy of 14TeV~\cite{CMSBsmumu} and assuming an integrated luminosity of 1fb$^{-1}$. 
The different sensitivities can be understood by comparing the different invariant mass resolution of the 
three experiments, which is one of the best discriminating variables of this analysis. 
The sensitivity scales roughly as the inverse square root of the invariant mass resolution. 
According to the Monte Carlo simulation the invariant mass resolutions for  LHCb, CMS and ATLAS are 
respectively 20MeV, 53MeV and 90MeV. \\
While LHCb has the best sensitivity for a given luminosity, 
ATLAS and CMS can take advantage of their capability of operating at much higher luminosities 
than those which will be used at LHCb. 
Finally it is worth noticing that the LHCb experiment has the possibility of discovering NP with 1fb$^{-1}$ 
if the branching ratio $B_s \rightarrow \mu^+ \mu^-$ is larger than $17 \times 10^{-9}$.

\Acknowledgements
The author would like to thank the of the CKM 2010 conference for hospitality and the excellent 
program.

\end{document}